\documentclass[12pt]{article}
\usepackage{amsmath,amsfonts,amsthm}
\usepackage{epsfig}
 
\newcommand{\charosculation}{\epsfig{file=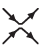, width=5mm}}
\newcommand{\charswitch}{\epsfig{file=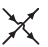, width=5mm}}

\newcommand{\figvertexconfig}{\epsfig{file=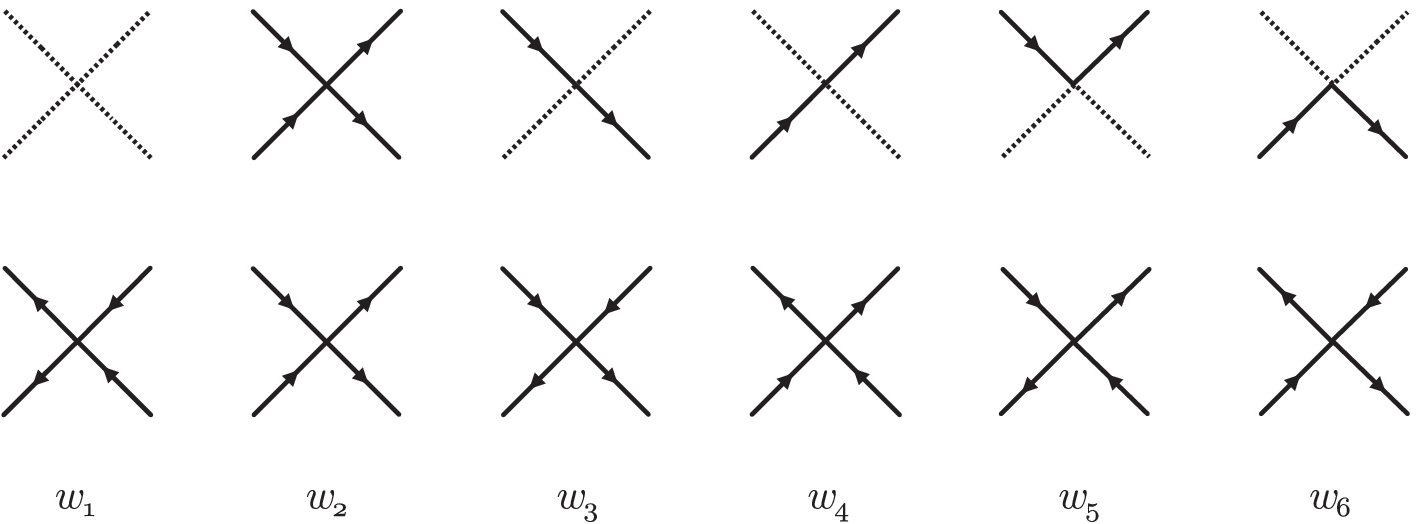,width=15cm}}
\newcommand{\figlattice}{\epsfig{file=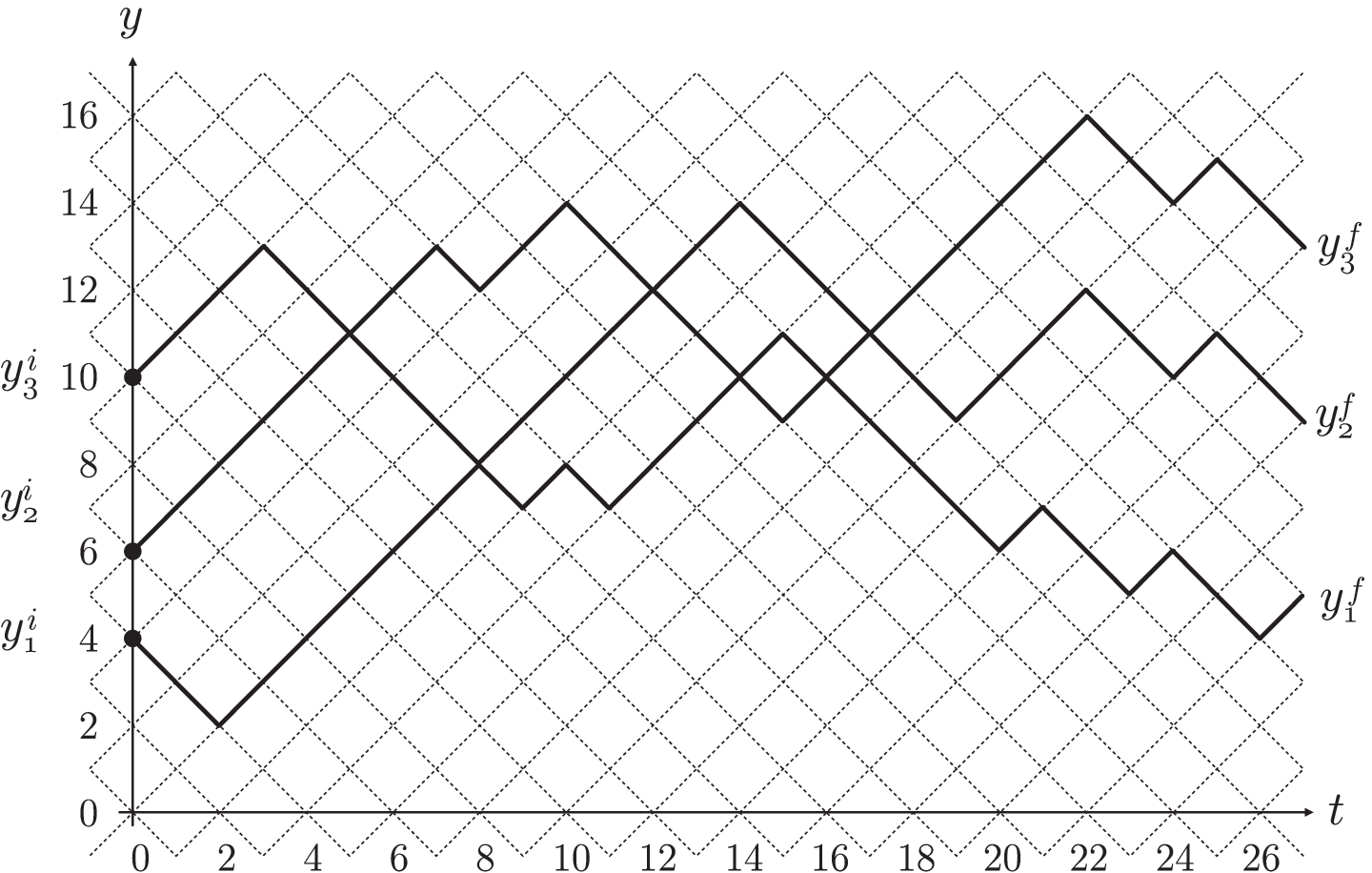,width=15cm}}

\newlength{\shift}
\newcommand{\pathfigA}{\raisebox{\shift}{\epsfig{file=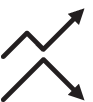,width=5mm}}}
\newcommand{\pathfigB}{\raisebox{\shift}{\epsfig{file=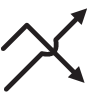,width=5mm}}}
\newcommand{\pathfigC}{\raisebox{\shift}{\epsfig{file=switch.fr8-3.eps,width=5mm}}}
\newcommand{\pathfigD}{\raisebox{\shift}{\epsfig{file=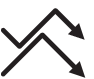,width=5mm}}}
\newcommand{\pathfigE}{\raisebox{\shift}{\epsfig{file=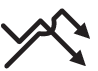,width=5mm}}}

\newcommand{\pto}{\!\rightarrow\!}

\newcommand{\sqlat}{\mathcal{L}}

\newcommand{\cI}{\mathcal{I}}
\newcommand{\cW}{\mathcal{W}}

\newcommand{\cE}{\mathcal{E}}

\newcommand{\cF}{\mathcal{F}}

\newcommand{\al}{\alpha}
\newcommand{\be}{\beta}
\newcommand{\si}{\sigma}

\newcommand{\ep}{\epsilon}
\newcommand{\Om}{\Omega}

\newcommand{\byi}{\mbf{y}^{i}}
\newcommand{\byf}{\mbf{y}^{f}}

\newcommand{\bX}{\mbf{X}}
\newcommand{\Omo}{ \Om^{s}}
\newcommand{\Omc}{\hat{\Om}^{s}}
\newcommand{\cP}{\mathcal{P}}
\newcommand{\Omni}{\Om^{n}}

\newcommand{\mbf}[1]{\mathbf{#1}}

\theoremstyle{plain}

\newtheorem*{thm}{Theorem [Free Fermion Walks]}

\theoremstyle{definition}

\newtheorem*{rem}{Remark}

\begin{document}
\pagenumbering{roman} \setcounter{page}{0} 
\title{A Combinatorial Interpretation of the Free Fermion
Condition of the Six-Vertex Model} 
\author{R. Brak and A. Owczarek\thanks{{\tt {\rm email:} r.brak,aleks@ms.unimelb.edu.au}},\\
Department of Mathematics and Statistics,\\
The University of Melbourne,\\
Parkville, Victoria 3052, Australia
}

\date{
\begin{center}
	10th March, 1999 
\end{center}
}
\maketitle 
\begin{abstract} 
The free fermion condition of the six-vertex model provides a 5 parameter 
sub-manifold on which the Bethe Ansatz equations for the wavenumbers that enter 
into the eigenfunctions of the transfer matrices of the model decouple, hence 
allowing explicit solutions.  Such conditions arose originally in early 
field-theoretic S-matrix approaches.  Here we provide a combinatorial 
explanation for the condition in terms of a generalised Gessel-Viennot 
involution.  By doing so we extend the use of the Gessel-Viennot theorem, 
originally devised for non-intersecting walks only, to a special weighted type 
of \emph{intersecting} walk, and hence express the partition function of $N$ 
such walks starting and finishing at fixed endpoints in terms of the
single walk partition functions.

\vspace{1cm} 

\end{abstract} 
\vfill
\newpage
\pagenumbering{arabic}

\section{Introduction} 

There has been a steady stream of interest in the statistical
mechanics of directed walk problems because of their connections to
the physics of polymers and domain walls \cite{fisher84}. Since the
popularisation of the field in the seminal article by Fisher
\cite{fisher84}, vicious walkers, in particular, also known as
non-intersecting walks, on two-dimensional (directed) lattices have
been the subject of much work
\cite{forrester89,forrester89b,forrester90,forrester91,brak98a,guttmann98a}. In
the field of combinatorics a very general methodology, useful for any
directed graph and based on a involution, has been developed by Gessel
and Viennot \cite{gessel85,gessel89} following the work of
Lindstr\"{o}m \cite{lindstrom73}, and Karlin and McGregor
\cite{karlin59}, which expresses the generating function of configurations of $N$
walks as the value of a determinant of single walk generating
functions. In the most general setup an arbitrary inhomogeneous weight
may be associated with each occupied edge of the lattice. 

\begin{figure}[ht!]
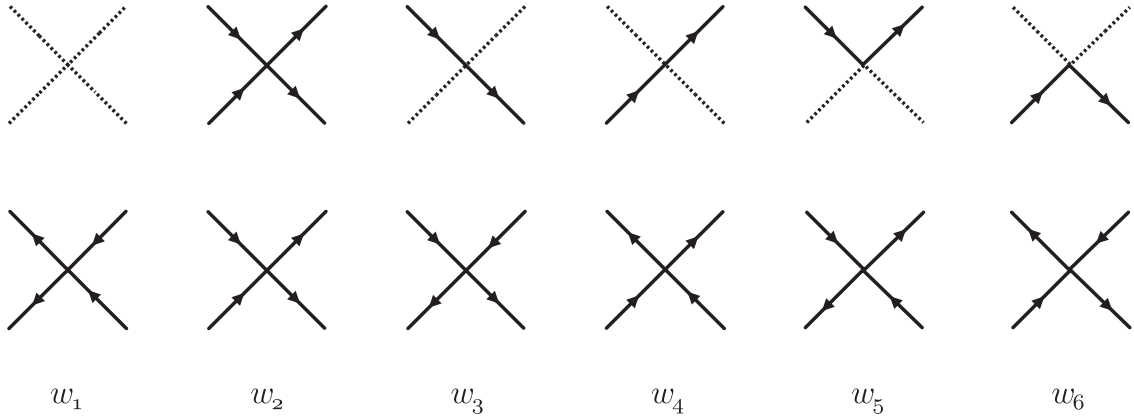

   \centering
\figvertexconfig
\caption{\it At the top are the six possible path configurations at a
vertex of the lattice. Below each of these is the associated arrow
configuration of the six-vertex model, while below that is the six
weights we associate with each of those configurations.}\label{fig:vertexconfig}
   \end{figure}

On the other hand it has been well known for a long time
\cite{wu68a,guttmann98a} that the square lattice six-vertex model can
be mapped onto a problem of interacting (intersecting) directed walks
on that lattice (see figure~\ref{fig:vertexconfig}). Let us call the
six weights of that model $\{w_1,w_2,w_3,w_4,w_5,w_6\}$: see
figure~\ref{fig:vertexconfig}. We can consider $w_1=1$ without loss of
generality. In order to calculate the partition function of fixed
numbers of walks one needs to consider a particular invariant
sub-space of the associated transfer matrix, the diagonalisation of
which involves the famous Bethe Ansatz trial solution: The Bethe
Ansatz is a guess for the eigenvectors of the transfer matrix and
involves a sum over a set of plane wave forms. For a problem of $N$
walks the Bethe Ansatz involves $N$ wavenumbers which are chosen from
the solutions of a set of $N$ non-linear coupled polynomial
equations. To calculate the walk partition function one needs to find,
and be able to sum over, all the eigenvalues and eigenvectors
explicitly.  

Recently \cite{brak98b} it has been shown that if one rather considers
the combinatorialist's problem of $N$ vicious walkers with weights
associated with edges, rather than with vertices, this can be solved
using the transfer matrix/Bethe Ansatz approach in a completely
rigorous fashion. Here the Bethe Ansatz equations for $N$ walks
decouple and the solution of the $N$ walk problem is given by the
Gessel-Viennot determinant of single walk generating functions. 
The walk problem using edge weights is equivalent to a \emph{restricted} vertex model 
(or visa versa)  where the vertex weights must satisfy the equations
\begin{align}
\label{eq:cond1}
w_3 w_4 &= w_5 w_6   \\
\intertext{and} 
\label{eq:cond2}
w_2&=0
\end{align}
The second condition \eqref{eq:cond2} merely expresses the fact
that the weight associated with the meeting of two walks is set to
zero since vicious walkers are being considered. The associated vertex
model with only this second condition necessarily holding is often
referred to as the five vertex model. 

  Of central importance here is that taken together the two conditions, 
  \eqref{eq:cond1} and \eqref{eq:cond2}, imply that the less restrictive 
  free-fermion condition of the six-vertex model, which occurs when the vertex 
  weights satisfy the equation
\begin{equation}
\label{eq:ffcond}
w_1 w_2 = w_5 w_6-w_3 w_4 
\end{equation}
    is then automatically satisfied.  Hence the edge weight model is equivalent 
    to the free-fermion case of the five-vertex model.  This is not surprising 
    since the free-fermion condition \eqref{eq:ffcond} is precisely the general 
    condition needed to achieve the decoupling of the Bethe Ansatz equations in 
    the solution of the six-vertex model.  
    
    This raises the question of whether the Gessel-Viennot methodology can be 
    adapted to the `free-fermion' case (i.e.\ equation \eqref{eq:ffcond} is 
    satisfied), of `six-vertex' or `osculating' walks -- here   the walks 
    \emph{are allowed to intersect} but not share edges i.e.\ site-only 
    intersecting.   


  In this paper we demonstrate that free-fermion osculating walks can indeed be 
  counted with a generalisation of the Gessel-Viennot methodology and we hence 
  explicitly calculate their generating function.  Because the Gessel-Viennot 
  involution is involved the result is again a determinant of single walk 
  generating functions.  We restrict our discussion to the square lattice but 
  the ideas can be easily generalised to any directed (acyclic) graph.
\begin{figure}[ht!]
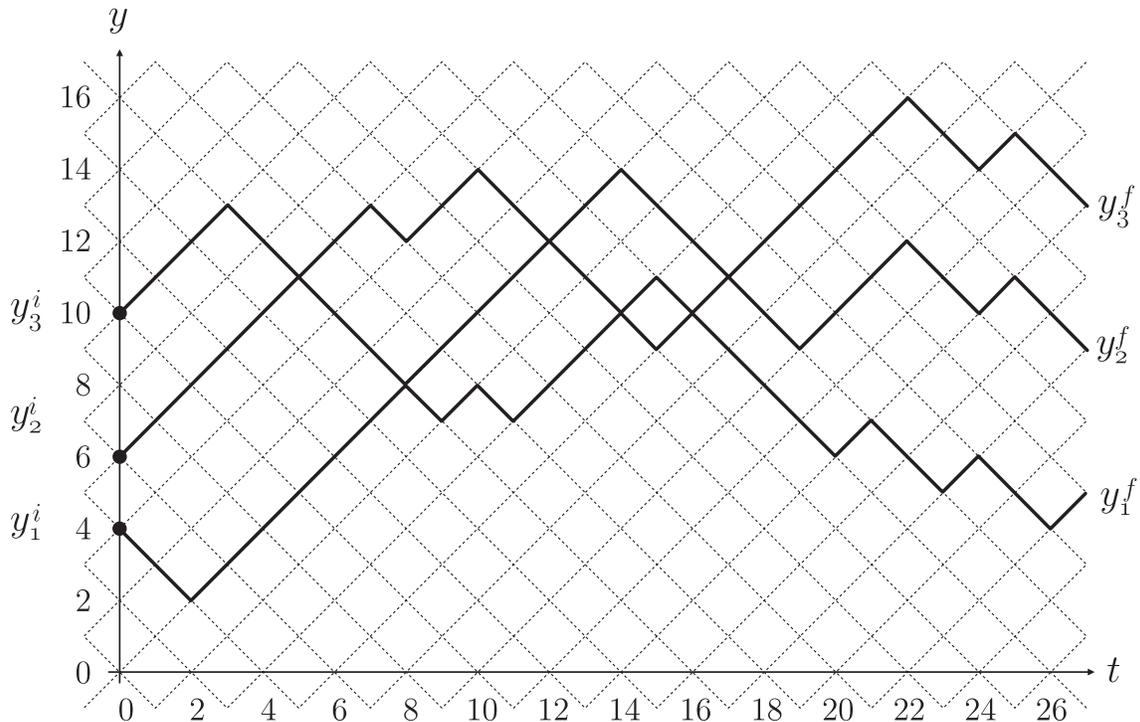

   \centering
  \figlattice
\caption{\it An example of three osculating paths.}\label{fig:lattice}
   \end{figure}
  In order to understand the combinatorial interpretation of the osculating 
  free-fermion walks we briefly discuss the case of non-intersecting walks for 
  those unfamiliar with the Gessel-Viennot method.  For the case of 
  \emph{non-intersecting} walks the Gessel-Viennot theorem gives the generating 
  function of $N$ such paths starting and ending at some fixed sets of sites as 
  the determinant of the generating functions of independent-single-walks 
  where the walks start and end at all permutations 
  of those endpoints.  That is, the generating function, 
  $\mathcal{N}^{(N)}_{t}({\bf y}^i\pto{\bf y}^f)$, for $N$ \emph{non}-intersecting walk 
  configurations where the $j^{th}$ walk starts at $y_j^i$ and arrives at 
  $y_j^f$ after $t$ steps is given by the following determinant:
\begin{equation}
                \mathcal{N}^{(N)}_{t}({\bf y}^i\pto {\bf y}^f) =
                \det || \mathcal{N}^{(1)}_{t}(y^i_\alpha \pto {y}^f_{\beta})
                ||_{\alpha,\beta=1\ldots N}.
\end{equation}
  where ${\bf y}^{i}=(y^{i}_1,\ldots,y^{i}_N)$ and ${\bf
  y}^{f}=(y^{f}_1,\ldots,y^{f}_N)$. Hence the generating function of
  $N$ walks is given as the signed sum over products of $N$ single
  walk generating functions where the sign of the contribution is
  signature of the permutation of the endpoints.  This signed sum is
  interpreted as a signed sum over the elements of a ``signed'' set,
  $\Om$, the elements of which are configurations of $N$, possibly
  intersecting, walks,
\begin{equation}
                \mathcal{F}^{(N)}_{t}({\bf y}^i\pto {\bf y}^f) =
                \sum_{\bX\in\Om}\ep_{\bX}\prod_{\al=1}^{N}
		\mathcal{M}^{(1)}_{t}(X_{\al}).
		\label{eq:ee}
\end{equation}
   where $X_{\al}$ is a single walk from $y^i_\alpha \pto
   {y}^f_{\beta}$, $\ep_{\bX}$ is the sign of the configuration and
   $\mathcal{M}^{(1)}_{t}(X_{\al})$ the weight of the a particular
   configuration of a \emph{single} walk -- see section \ref{sec:defs}
   for more precise definitions of the terms referred to in this
   section.  Note that the $N$ single walks in elements of $\Om$ may
   be edge as well as site intersecting, as they are completely
   independent of each other.
   
   The method introduced by Gessel and Viennot \cite{gessel85,gessel89} shows 
   that pairs of opposite signed terms of \eqref{eq:ee} corresponding to 
   configurations with at least one intersection cancel leaving only  
   positively signed terms corresponding to non-intersecting configurations -- 
   precisely the sum required to give $\mathcal{N}^{(N)}_{t}({\bf y}^i\pto {\bf 
   y}^f)$. This pairing rule is an ``involution''.
  
   In order to interpret the free-fermion condition for osculating or six-vertex 
   walks we apply the same type of methodology but now require an extended 
   pairing rule, one for which the pairs of terms in the signed sum do 
   \emph{not} necessarily cancel out.  We now consider three different disjoint 
   subsets of $\Om$: $\Om^{n}$ the subset containing configurations that do not intersect 
   at all, $\Om^{s}$ the subset containing configurations that only share sites (and not 
   edges) and $\Om^{e}$ the subset containing configurations that share at least one edge.
   
   The involution now pairs oppositely signed terms from $\Om^{e}$
   which cancel out, but the pairs of oppositely signed terms from the
   subset $\Om^{s}$ have different weights and \emph{do not} cancel
   out, rather the weight difference is precisely given by the
   free-fermion condition \eqref{eq:ffcond}. This idea is expressed
   schematically by the case for (example) paths in the subset $\Om^{s}$
   as
\begin{align}
    \cP(\pathfigA)-\cP(\pathfigB)&=\cW(\pathfigC)\\
   \intertext{and for walks in the subset $\Om^{e}$ as}
   \cP(\pathfigD)-\cP(\pathfigE)&=0
\end{align}
  where $\cP(X)$ is the \emph{product} of the vertex weights of the
  two single paths of the configuration $X$, whilst $\cW(X)$ is the
  vertex weight of the paths, $X$, \emph{taken as a whole}.

  Thus the free fermion condition arises as a natural consequence of summing 
  over a signed set of $N$-walk configurations each of whose weight is a 
  \emph{product} of single path vertex weights.  The intersecting configurations 
  in the signed set which do not cancel combine to create the correct vertex 
  weight for the $N$-walk configuration.

\section{Free Fermion walk generating function}
\label{sec:defs}
\paragraph{Definitions and Notations}

We will consider walks on the directed square lattice, $\sqlat$ rotated through 
$45^\circ$.  Each vertex of the lattice is labelled by the ``time'' coordinate, 
$t$ and a height $y$ and represented by the pair $(t,y)$ or function $y(t)$ -- 
see figure \ref{fig:lattice}.  An $N$-\emph{vertex} is an $N$-tuple of vertices of 
$\sqlat$ each of has the same time coordinate and is represented by the $N$-tuple 
of their height coordinates, $\mbf{y}=(y_{1},\ldots,y_{N})$.  An $N$-vertex 
$\mbf{y}$ is \emph{non-intersecting} if $y_{\be}\ne y_{\al}$, $\al,\be=1,\ldots, 
N$, $\al\ne\be$.  A walk of length $t$, $X(y_{\al},y_{\be}) = y(0)y(1)y(2)\ldots 
y(t)$, where $y(0)=y_{\al}$ and $y(t)=y_{\be}$, is a sequence of $t$ adjacent 
edges from vertex $ y(0)$ to vertex $ y(t)$.  An $N$-walk, 
$\mbf{X}(\mbf{y}^{i},\mbf{y}^{f})$ is an $N$-tuple of equal length walks, 
$(X_{1}(y^{i}_{1},y^{f}_{1}), \ldots,X_{N}(y^{i}_{N},y^{f}_{N}))$, with 
$\mbf{y}^{i}$ and $\mbf{y}^{f}$ non-intersecting.

An \emph{osculation} between walks $X_{\al}$ and $X_{\be}$ at time $s$ occurs if 
$y_{\al}(s)=y_{\be}(s)$, and either $y_{\al}(s-1)<y_{\be}(s-1)$ and 
$y_{\al}(s+1)<y_{\be}(s+1)$ or $y_{\al}(s-1)>y_{\be}(s-1)$ and 
$y_{\al}(s+1)>y_{\be}(s+1)$ occur i.e the configuration \charosculation\ appears.  
Paths $X_{\al}$ and $X_{\be}$ \emph{switch} at time $s$ if 
$y_{\al}(s)=y_{\be}(s)$, and either $y_{\al}(s-1)<y_{\be}(s-1)$ and 
$y_{\al}(s+1)>y_{\be}(s+1)$ or  $y_{\al}(s-1)> y_{\be}(s-1)$ and 
$y_{\al}(s+1)<y_{\be}(s+1)$ occur i.e the configuration \charswitch\ appears.  Since 
the initial and final $N$-vertices are non-intersecting we do not define any 
osculations or switches associated with these vertices.  A pair of walks is 
\emph{osculating} if all common vertices are osculations -- see figure 
\ref{fig:lattice}.  An $N$-walk $\mbf{X}$ is osculating if all the   vertices in 
common with any of the walks form osculations.

Let $X_{\al}=y_{\al}(0)y_{\al}(1)y_{\al}(2)\ldots y_{\al}(t)$ and
$X_{\be}=y_{\be}(0)y_{\be}(1)y_{\be}(2)\ldots y_{\be}(t)$ intersect at
time $s$ i.e $y_{\al}(s)=y_{\be}(s)$, then the pair $X_{\al}$ and
$X_{\be}$ are \emph{exchanged} at time $s$ if
\begin{align}
    X_{\al}\to X'_{\al}&=y_{\al}(0) \ldots 
	y_{\al}(s)y_{\be}(s+1)\ldots
	y_{\be}(t)\notag\\
X_{\be}\to X'_{\be}&=y_{\be}(0) \ldots 
	y_{\be}(s)y_{\al}(s+1)\ldots
	y_{\al}(t)
\label{eq:exchange}
\end{align}

Define $[N]= \{1,\ldots,N\}$.  Let $P_{N}$ be the set of permutations
of $[N]$, then for $\si=(\si_{1},\ldots,\si_{N})\in P_{N}$, and
$\mbf{y}=(y_{1},\ldots,y_{N})$ an $N$-vertex, then
$\si(\mbf{y})=(y_{\si_{1}},\ldots,y_{\si_{N}})$.  The signature of a
permutation is denoted $\ep_{\si}$.

We will associate weights with  the walks on the lattice as
follows.  Associate a set of six vertex weights,
\begin{equation}
    \mathcal{V}(v)=\{ w_{1}(v),\ldots, w_{6}(v)\},
\end{equation}
with each vertex $v\in\sqlat$.  Without loss of generality we only
consider the situation where \emph{five} of the weights are not equal
to one.  The vertex weight, $\cW(\bX)$ of a particular $N$-walk, $\bX$,
is the product of the vertex weights of all the vertices of $\sqlat$
that are traversed by the walks of $N$-walk.  For each vertex traversed
by the $N$-walk, only one of the six possible vertex weights associated
with the vertex of $\sqlat $ is used.  Which of the six possible
weight used depends on which of the four edges adjacent to the
particular vertex are traversed by the $N$-walk as illustrated in figure
\ref{fig:vertexconfig}.  No vertex weights are associated with the initial and 
final $N$-vertices of the $N$-walk.

\begin{rem}
Note that with the above definition of osculating walks the set 
includes non-intersecting walks also.
\end{rem}
\begin{rem} The vertex weight $\cW(\bX)$ is only defined for
osculating paths., i.e if the paths of the $N$-walk do not cross or if
$\bX$ contains only one path.
\end{rem}
\begin{rem}
This way of associating weights with the walks is a generalisation of the 
situation dealt with by the Gessel-Viennot theorem where the weights are 
associated with the edges of the graph and are independent of the $N$-walk 
configuration.  The problem of finding the generating function for osculating 
walks with arbitrary weights requires a rather complicated involution 
\cite{brak99x}, however in the special case where the vertex weights satisfy the 
free fermion equation the osculating walk generating function can be expressed 
as a determinant, as we demonstrate here.
\end{rem}
  

\begin{thm} 
Let  $\Om^{*}$ be the set of all osculating $N$-walks of length $t$ 
starting at $\byi$ and ending at $\byf$ with $y^{i}_{\al}<y^{i}_{\al+1}$ and 
$y^{f}_{\al}<y^{f}_{\al+1}$, $\al\in[N-1]$.
If, for $v\in\sqlat$, the vertex weights satisfy
\begin{equation}
w_5(v)w_6(v)-w_3(v) w_4(v)=w_2(v)
\label{eq:ff}
\end{equation}
then the   osculating lattice walk generating function,
\begin{equation}
     \cF_{t}^{(N)}(\byi\to\byf)=\sum_{\bX
\in\Om^{*}}\cW(\bX)
\end{equation} 
   where $\cW(\bX)$ is the vertex weight of the $N$-walk $\bX$, 
is given by
\begin{equation}
     \cF_{t}^{(N)}(\byi\to\byf)=\text{Det}|| \cF_{t}^{(1)}(y^{i}_{\al}\to 
y^{f}_{\be})||_{\alpha,\beta=1\ldots N}
\end{equation} 
where $\cF_{t}^{(1)}(y^{i}_{\al}\to y^{f}_{\be})$ is the
generating function for a \emph{single} lattice walk from
$y^{i}_{\al}\to y^{f}_{\be}$.
\end{thm}
The theorem is proved by an extension of the Gessel-Viennot involution
which \emph{does not} preserve the $N$-walk weight.

\begin{proof}
  
Consider the signed set $\Om=\Om^{+}\cup\Om^{-}$,
$\Om^{+}\cap\Om^{-}=\phi$, where $\phi$ is the empty set.  The
positive and negative sets are
\begin{align}
    \Om^{+}&=\{\bX(\mbf{y}^{i},\si(\mbf{y}^{f})) | \text{$\si\in P_{N}$ and  
$\ep_{\si}=+1$} \}\\
    \Om^{-}&=\{\bX(\mbf{y}^{i},\si(\mbf{y}^{f}))  | \text{$\si\in P_{N}$ and  
$\ep_{\si}=-1$}  \}
\end{align} 
Let, $\bX\in\Om$, then the sign of $\bX $ is defined as
\begin{equation}
    \ep_{\bX}=\left\{
    		\begin{array}{ll}
		    +1 & \text{if $\bX\in\Om^{+}$}\\
		    -1 & \text{if $\bX\in\Om^{-}$}
		    \end{array}
    		\right.
		\label{eq:sign}
\end{equation}    
For the $N$-walks of $\Om$ we do \emph{not} use the vertex weight of the
$N$-walk as a whole, but rather define a ``product'' weight, $\cP(\bX)$
in terms of its individual walks.  In particular,
\begin{equation}
      \cP(\bX)=\prod_{X_{\al}\in\bX}\cW(X_{\al})
      \label{eq:prodform}
\end{equation}
  
We  construct a sign reversing involution\footnote{A sign reversing involution, $\psi$ is a permutation of
$\Om$ such that $\psi^{2}=\text{Identity}$ and it has the property
that whenever $\psi(\bX)\ne\bX$, then $\bX\in \Om^{+}$, if and only if
$\psi(\bX)\in \Om^{-}$.}, $\psi$
on $\Om$. 
  
Let $\bX=(X_{1},\ldots,X_{N}) \in\Om$.  The involution is an
extension of the Gessel-Viennot involution and splits into three
cases,
\begin{enumerate}
\item \emph{No intersections} ($\bX\in\Omni$). If none of the walks of $\bX$ intersect, 
then $\psi(\bX)=\bX$.
  
\item \emph{Edge intersections} ($\bX\in\Om^{e}$). If any of the walks $X_{\al}\in\bX$ has an 
edge of $\sqlat$ in common with another walk,   $X_{\be}\in\bX$ then
let $\al$ be the least integer for which a walk $X_{\al}$ shares an edge
with another walk $X_{\be}$.  Of all the vertices adjacent to the
edges in common with $X_{\al}$ and $X_{\be}$ choose the one with the
smallest $t$ coordinate and denote it by $v$, then $\bX'=\psi(\bX)$ is
defined as the $N$-walk obtained by exchanging $X_{\al}$ and $X_{\be}$
at $v$.
  
\item \emph{Vertex only intersections} ($\bX\in\Om^{s}$). If any of the walks of $\bX$ share 
vertices and none share edges, then let $\al$ be the least integer for
which a walk $X_{\al}$ intersects another walk $X_{\be}$.  Of all the
vertices in common with $X_{\al}$ and $X_{\be}$ choose the one with
the smallest $t$ coordinate and denote it by $v$, then
$\bX'=\psi(\bX)$ is defined as the $N$-walk obtained by exchanging
$X_{\al}$ and $X_{\be}$ at $v$.
\end{enumerate}
Call the vertex, $v$ at which the involution exchange takes place, the 
``involution'' vertex.  The difference in the product weight of $\bX$ and 
$\psi(\bX)$ is then
\begin{equation}
	\cP(\bX )-\cP( \psi(\bX))=\left\{\begin{array}{ll}
		 \bigl(w_{5}(v)w_{6}(v)-w_{3}(v)w_{4}(v)\bigr)\,\cP_{\{v\}}(   \bX )
		 &\text{if $v$ is an   osculation}\\
		 &\\  
		 \bigl(w_{3}(v)w_{4}(v)-w_{5}(v)w_{6}(v)\bigr)\,\cP_{\{v\}}(   \bX )
		 &\text{if $v$ is a switch}\\
		 &\\
	 0&\text{otherwise}
	 \end{array}\right.
    \label{eq:ffws}
\end{equation}
where $\cP_{\{v\}}( \bX )$ is the product weight of $ \bX $ with the
contribution of the weight associated with $v$ divided out.  
  
Since $\Omo\,\subset\Om$ is defined as the set of $N$-walks for which the
involution vertex $v$ exists and arises from the ``vertex only
intersections'' case of $\psi$, this means that all the $N$-walks
in $\Omo$ have walks which only intersect at vertices -- there are no
shared edges.  Define two $N$-walks to be related, $\bX\sim\bX'$ iff
$\bX$ can be obtained from $\bX'$ by the interchange of any number of
osculations with switches (or visa versa).  This relation is easily
seen to be an equivalence relation and hence partitions $\Omo$ into
disjoint subsets, $\Omc_{\al}$, $\al\in\cI$ where $\cI$ is some index
set for the partitions.  Define the canonical element, $\bX^{c_\al}$
of each partition, $\Omc_{\al}$, as the $N$-walk for which all the
intersections are osculations.  Let $\cE(\bX^{c_\al})$ be the set of
vertices of $\bX^{c_\al}\in\Omc_{\al}$ in common with at least two 
walks of $\bX^{c_\al}$.  Note, the cardinality of $\Omc_{\al}$ is
$2^{|\cE(\bX^{c_\al})|}$.
  
For the
$N$-walks in $\Omc_{\al}$ we have,
\begin{align}
      \sum_{\bX\in\Omc_{\al}}\ep_{\bX}\cP(\bX)=
         \cP_{\cE}(\bX^{c_\al})\prod_{z\in\cE(\bX^{c_\al})}
      \bigl(w_{5}(z)w_{6}(z)-w_{3}(z)w_{4}(z)\bigr) 
\label{eq:congen}
\end{align}  
where $\cP_{\cE}(\bX^{c_\al})$ is the product weight of
$\bX^{c_\al}$ with all the vertex weights associated with the vertices
in $\cE(\bX^{c_\al})$ divided out.  This follows since
$\prod_{z\in\cE(\bX^{c_\al})}
\bigl(w_{5}(z)w_{6}(z)-w_{3}(z)w_{4}(z)\bigr)$ allows for each vertex
in $\cE(\bX^{c_\al})$ to be a switch (i.e.\ weight $w_{3}w_{4}$) or an
osculation (i.e.\ weight $w_{5}w_{6}$).  The sign, $\ep_{\bX}$ is
correctly obtained since it is just $-1$ to the number of occurrences
of a switch i.e.\ the number of factors of $-w_{3}w_{4}$. Thus we 
have the following:
\begin{align}
    \text{Det}||& \cP(y^{i}_{\al}\to y^{f}_{\be})||_{\alpha,\beta=1\ldots N}
       =\sum_{\bX\in\Om}\ep_{\bX} \cP(\bX)\\
       &=\sum_{\bX\in\Om - \Omo}\ep_{\bX} 
       \cP(\bX)+\sum_{\bX\in\Omo}\ep_{\bX} \cP(\bX)\\
   \intertext{and since, by \eqref{eq:ffws} the $N$-walks in $\Om -  \Omo $ ($= \Omni \cup \Om^{e}$) with any 
   intersections   
   cancel in pairs we get}
   	&=\sum_{\bX\in\Omni}\cW(\bX)+\sum_{\bX\in\Omo}\ep_{\bX} \cP(\bX)\\
       &=\sum_{\bX\in\Omni}\cW(\bX)+\sum_{\al\in\cI}\,\sum_{\bX\in\Omc_{\al}}\ep_{\bX} 
       \cP(\bX)\\
\intertext{where $\Omni\subset\Om$ is the set of non-intersecting $N$-walks, using 
\eqref{eq:congen} gives,} 
     &=\sum_{\bX\in\Omni}\cW(\bX)+\sum_{\al\in\cI} 
\cP_{\cE}(\bX^{c_\al})\prod_{z\in\cE(\bX^{c_\al})} 
\bigl(w_{5}(z)w_{6}(z)-w_{3}(z)w_{4}(z)\bigr)\\
   \intertext{now, using the free fermion relation, \eqref{eq:ff} we get}
      &=\sum_{\bX\in\Omni}\cW(\bX)+\sum_{\al\in\cI} 
        \cP_{\cE}(\bX^{c_\al})\prod_{z\in\cE(\bX^{c_\al})}w_{2}(z)\\
	 &=\sum_{\bX\in\Omni}\cW(\bX)+\sum_{\al\in\cI} 
        \cW(\bX^{c_\al})\\
   \intertext{and since all the canonical $N$-walks, $\bX^{c_\al}$ are osculating 
   we get}
	&=\sum_{\bX 
\in\Om^{*}}\cW(\bX)
\end{align}  
 as required. 
 
\end{proof}

\section*{Acknowledgements}
Financial support from the Australian Research Council is gratefully
acknowledged by the authors.

\bibliography{annals-comb}
\bibliographystyle{aip}

\end{document}